\documentclass{aastex}
\usepackage{spr-astr-addons}
\usepackage{graphicx}
\shorttitle{Electron-positron annihilation lines}
\begin{document}
\title{Electron-positron Annihilation Lines and 
Decaying Sterile Neutrinos}
\author{M.~H.~Chan and M.~-C.~Chu}
\affil{Department of Physics and Institute of Theoretical Physics,
\\ The Chinese University of Hong Kong,
\\  Shatin, New Territories, Hong Kong, China}
\email{mhchan@phy.cuhk.edu.hk, mcchu@phy.cuhk.edu.hk}
\begin{abstract}
If massive sterile neutrinos exist, their decays into photons and/or 
electron-positron pairs may give rise to observable consequences. We 
consider the possibility that MeV sterile neutrino decays lead to the 
diffuse positron annihilation line in the Milky Way center, and we thus 
obtain bounds on the sterile neutrino decay rate $\Gamma_e \ge 10^{-28}$ 
s$^{-1}$ from 
relevant astrophysical/cosmological data. Also, we expect
a soft gamma flux of $1.2 \times 10^{-4}-9.7 \times 10^{-4}$ ph 
cm$^{-2}$ s$^{-1}$ from the Milky Way center which shows up as a small MeV 
bump in the background photon spectrum. Furthermore, we estimate the flux 
of active neutrinos produced by sterile neutrino decays to be $0.02-0.1$ 
cm$^{-2}$ s$^{-1}$ passing through the earth. 
\end{abstract} 

\section{Introduction}
Understanding the nature of dark matter remains a fundamental problem in 
astrophysics and cosmology. Since the discovery of neutrinos' non-zero 
rest mass \citep{Fukuda,Bilenky}, the possibility 
that neutrinos contribute to cosmological dark matter has become a hot 
topic again. In particular, the 
sterile neutrinos belong to a class of candidate 
dark matter particles with no 
standard model interaction. Although the recent 
MiniBooNE data challenges the LSND result that 
suggests the existence of eV scale sterile neutrinos \citep{Aguilar}, more 
massive sterile neutrinos (eg. keV, MeV) may still exist. The fact that 
active 
neutrinos have mass implies that right-handed neutrinos should exist 
which may indeed be massive sterile neutrinos. The existence of the 
sterile neutrinos has been invoked to explain many phenomena such as 
missing mass \citep{Dodelson,Fuller} and the high temperature of the hot 
gas in 
Milky Way and clusters \citep{Chan,Chan2}. Therefore, it is worthwhile to 
discuss observational consequences if massive sterile 
neutrinos exist, 
which may decay into light neutrinos, positron-electron pairs and 
photons. In this article, we consider the possibility that sterile 
neutrino decays give rise to the 511 keV lines in Milky Way 
and thus obtain bounds on the mass $m_s$ and
total decay rate $\Gamma$ of the sterile neutrinos using relevant
observational data.

\section{511 keV photon flux}
The bright 511 keV annihilation line from Milky Way has
been observed for a few decades \citep{Leventhal,Knodlseder}, and its 
origin has been much debated. Recent values of the
511 keV photon flux from the bulge and disk are $(1.05 \pm 0.06) \times 
10^{-3}$~ph~cm$^{-2}$~s$^{-1}$ and $(0.7 \pm 0.4) \times 
10^{-3}$~ph~cm$^{-2}$~s$^{-1}$ respectively \citep{Knodlseder}. Assuming a 
positronium fraction of $f_p=0.93$, one can translate these intensities to 
annihilation rates of 
$(1.5 \pm 0.1) \times 10^{43}$~s$^{-1}$ and $(0.3 \pm 0.2) \times 
10^{43}$~s$^{-1}$ respectively \citep{Knodlseder}. The annihilation rate 
in the 
bulge is several times larger than that in the disk. The source of 
positrons in the disk can be explained by the decay of $^{26}$Al. Using 
a disk model, the photon flux is calculated to be $5 \times 
10^{-4}$~ph~cm$^{-2}$~s$^{-1}$, which can account for $60-100 \%$ of the 
disk emission 
\citep{Knodlseder}. However, the origin of the bulge source is still 
an open 
question in astrophysics. Recent 
observation by INTEGRAL/IBIS indicates that the upper limit of photon flux 
for resolved single point sources is $1.6 \times 
10^{-4}$~ph~cm$^{-2}$~s$^{-1}$ \citep{Cesare}, which means that the 511 
keV annihilation line comes from mainly diffuse sources rather than point 
sources.

There have been many models trying to explain the 511 keV 
annihilation line of bulge emission. Potential sources include neutron 
stars or 
blackholes \citep{Lingenfelter}, supernova remnants \citep{Dermer}, 
Wolf-Rayet stars 
\citep{Ramaty}, pulsar wind \citep{Chi,Wang}, and Gamma Ray Bursts 
\citep{Lingenfelter2}. None of these 
can provide a satisfactory explanation as they are mainly 
point sources. \citet{Boehm} proposed the annihilation of dark matter
as a diffuse source. Recently, \citet{Picciotto} and 
\citet{Khalil} suggested that heavy sterile neutrinos with $m_s 
\ge 1$ MeV can be a diffuse source of the 511 keV photon flux. In the 
following, we extend the idea from these two papers and obtain
bounds on the decay rate of sterile neutrinos 
from observational data of the 511 keV annihilation line. 

\subsection{Decay of Sterile Neutrinos}
A sterile neutrino $\nu_s$ can decay into an electron-positron pair, 
photons 
and lighter neutrinos $\nu$ through different channels. The major channel 
is $\nu_s 
\rightarrow 3 \nu$ with decay rate \citep{Barger}
\begin{equation}
\Gamma_{3 \nu}= \frac{G_F^2}{384 \pi^3} \sin^22 \theta m_s^5=1.77 \times 
10^{-20} \sin^22 \theta \left( \frac{m_s}{\rm 1~keV} \right)^5~ \rm 
s^{-1},
\end{equation}
where $G_F$ and $\theta$ are the Fermi constant and mixing angle of 
sterile 
neutrino with active neutrinos respectively. The radiative channel is 
$\nu_s \rightarrow \nu + 
\gamma$ with decay rate \citep{Barger}
\begin{equation}
\Gamma_{\gamma}= \frac{9 \alpha G_F^2}{1024 \pi^4} \sin^22 \theta m_s^5= 
1.38 \times 10^{-22} \sin^22 \theta \left ( \frac{m_s}{\rm 1~keV} 
\right)^5~ \rm s^{-1},
\end{equation}
where $\alpha$ is the fine structure constant. An electron-positron pair 
is produced through $\nu_s \rightarrow \nu 
+e^{+}+e^{-}$ with decay rate \citep{Picciotto}
\begin{equation}
\Gamma_e= \frac{G_F^2}{384 \pi^3} \sin^22 \theta m_s^5 \left( 
\frac{|V|^2}{2}+ \frac{1}{8} \right)= \Gamma_{3 \nu} \left( 
\frac{|V|^2}{2}+ \frac{1}{8} \right),
\end{equation}
where $|V|<1$ is a parameter. Therefore, the total decay rate is 
\begin{equation}
\Gamma= \Gamma_{3 \nu}+ \Gamma_{\gamma}+ \Gamma_e \approx \Gamma_{3 \nu}+ 
\Gamma_e= \Gamma_e \left( \frac{4|V|^2+9}{4|V|^2+1} \right).
\end{equation}

\subsection{Positron channel}
The positrons produced will be slowed down due to ionization losses.  The 
power loss is approximately given by \citep{Longair}
\begin{equation}
\frac{dE}{dt} \sim -2 \times 10^{-10} \left( \frac{n}{{\rm 1~cm^{-3}}} 
\right) (\ln \gamma +6.6) \rm ~eV/s,
\end{equation}
where $\gamma$ is the Lorentz factor of the positron and $n$ is the 
average number 
density of electrons in the galactic bulge. The stopping distance $d$ for 
1 MeV positrons in this process is about $10^{24}$ cm. Also we should 
consider the magnetic field in the
Milky Way. The Larmor radius of positrons with energy $E_{e^+}$ is given 
by
\begin{equation}
r= \frac{E_{e^+}}{eB}=10^{13} \left( \frac{E_{e^+}}{10^{4}~\rm MeV} 
\right) \left( \frac{B}{10^{-5}~ \rm G} \right) \rm cm.
\end{equation}
The magnetic field strength in the Milky Way is about $B=10^{-5}$ G.
Therefore, for a 1 MeV positron, the Larmor radius is about $10^9$ cm. 
The stopping distance for the simple random walk 
of a positron, the distance that a positron is confined, is 
about $\sqrt{rd} \sim 1$ pc or less 
\citep{Boehm}, which is much shorter than the mean free path of the 
$e^{\pm}$ annihilation:
\begin{equation}
\bar{l}_{e^{\pm}}= \frac{1}{n \sigma_a} \sim \rm 30~ kpc,
\end{equation}
where
\begin{equation}
\sigma_a= \frac{\pi e^2}{m_ec^2( \gamma +1)} \left[ 
\frac{\gamma^2+4 \gamma +1}{\gamma^2-1} \ln (\gamma + \sqrt{ \gamma^2-1})- 
\frac{\gamma+3}{\sqrt{ \gamma^2-1}} \right]
\end{equation}
is the cross section of electron-positron annihilation \citep{Heitler}.
Therefore, the positrons will become non-relativistic before annihilation. 
However, the rate for a positron to annihilate with an electron in 
the diffuse region of Milky Way is
\begin{equation}
P \sim n \sigma_av_e \sim 10^{-18} \rm ~ s^{-1},
\end{equation}
where $n \approx 0.1$~cm$^{-3}$ \citep{Muno} and $v_e \approx 
10^7$~cm~s$^{-1}$ is the mean speed of electrons in Milky Way 
\citep{Marconi}. In order to 
produce $10^{43}$~s$^{-1}$ $e^{\pm}$ annihilations, there must exist a 
large positron cloud with $10^{61}$ positrons in the Milky Way, and the 
initial production rate should be much greater than the annihilation rate. 

Suppose a sterile neutrino halo is formed and the positron production rate 
is much higher than the 
annihilation rate during the galaxy formation due to the small $n$ in the 
protogalaxy. The positrons 
will accumulate in the protogalaxy. The rate of change in the positron 
number density $n_{e^+}$ is given by
\begin{equation}
\dot{n}_{e^+}=n_s(t) \Gamma_e-n_{e^+}n \sigma_av_e,
\end{equation}
where $n_s(t)$ is the number density of sterile neutrinos in the Milky Way 
at time $t$. Since 
we have $n_s(t)=n_{s0}e^{-\Gamma t}$, where $n_{s0}$ is the initial 
number density of sterile neutrinos, the solution of 
Eq.~(10) is 
\begin{equation}
n_{e^+}=e^{-n \sigma_av_et} \left( \int^t n_{s0} \Gamma_e e^{(n 
\sigma_av_e- \Gamma)t'}dt'+C \right),
\end{equation}
where $C$ is a constant. After a long time, assuming an equilibrium is 
established at present time $t_0$, we have $n_s(t_0) \Gamma_e= n_{e^+} 
n \sigma_av_e$. The total
annihilation rate in the bulge is given by
\begin{equation}
A_{\rm bulge} \approx \int_0^{R_B} 4 \pi r^2 n_s(t_0) \Gamma_e dr,
\end{equation}
where $R_B$ is the bulge radius, which is assumed to be $2.40-3.71$ kpc in 
the model used by \citet{Knodlseder}. Similarly, the annihilation rate in 
the disk is given by
\begin{equation}
A_{\rm disk} \approx \int_{R_B}^{R_D} 4 \pi rh n_s(t_0) \Gamma_e dr,
\end{equation}
where $R_D$ and $h$ are the radius and half of the thickness of the disk 
respectively. In 
the disk models used in \citet{Knodlseder}, the maximum $R_D$ is 15 kpc, 
and the scale heights of young and old disk models are 70 pc and 200 pc 
respectively. Here we assume that the sterile neutrino profile follows the 
dark matter profile, which can be modelled by the isothermal $n(s)=n_0
r^{-2}$ or NFW profile \citep{Navarro}:
\begin{equation}
n_s= \frac{n_0'}{r/a(1+r/a)^2},
\end{equation}
where $a$ and $n_0'$ are parameters in the NFW profile. In the isothermal 
profile, the ratio of the annihilation rate in the bulge to that in the 
disk is
\begin{equation}
\frac{A_{\rm bulge}}{A_{\rm disk}}= \frac{R_B}{h \ln(R_D/R_B)} 
\approx 6-13.
\end{equation}
In the NFW profile, the ratio is
\begin{equation}
\frac{A_{\rm bulge}}{A_{\rm disk}}= \frac{a}{h} \left[ \ln \left( 
\frac{R_B}{a}+1 \right)- \frac{R_B/a}{R_B/a+1} \right]\left( 
\frac{1}{R_B/a+1}- \frac{1}{R_D/a+1} \right)^{-1} \approx 1-4.
\end{equation}
Since over $60\%$ of disk emission can be explained 
by the decay of $^{26}$Al, the lower bound of the ratio of the diffuse 
emission should be about 7. Therefore, the isothermal profile 
agrees better 
with the observed ratio. Furthermore, since $A_{\rm bulge}=(1.5 
\pm 0.1) \times 10^{43}$~s$^{-1}$, from Eq.~(12), we have 
$n_0 \Gamma_e \sim 10^{22}$~m$^{-1}$~s$^{-1}$. The upper limit of 
central mass density in isothermal model constrains  
$m_sn_0 \le 5.5 \times 10^{19}$~kg~m$^{-1}$. For $m_s \ge 1$ MeV, 
we get $n_0 \le 3 \times 10^{49}$~m$^{-1}$. Therefore, $\Gamma \ge 3 
\times 10^{-28}$~s$^{-1}$. 
Similarly, for NFW model, we have $n_0' \Gamma_e \sim 4 \times 
10^{-19}$~m$^{-3}$~s$^{-1}$ and $m_sn_0' \le 3.2 \times 
10^{-22}$~kg~m$^{-3}$. For 
$m_s \ge 1$ MeV, we get $n_0' \le 2 \times 10^8$~m$^{-3}$ and $\Gamma_e 
\ge 2 \times 10^{-27}$~s$^{-1}$. If $\Gamma_e \sim 10^{-28}$~s$^{-1}$ and 
$m_s \sim 1$~MeV, we can get $\sin^22 \theta \sim 10^{-24}$, which is 
consistent with the diffuse X-ray background constraint \citep{Boyarsky}.

\subsection{Radiative channel}
There exists another radiative decay channel which 
gives a photon flux $\Phi_{\gamma}$ with energy $m_s/2$:
\begin{equation}
\Phi_{\gamma}= \int_{\rm line~of~sight} n_s \Gamma ds.
\end{equation}
The branching ratio is given by \citep{Picciotto}
\begin{equation}
\frac{\Phi_{\gamma}}{\Phi_{e^{\pm}}}= \frac{0.031}{4|V|^2+1},
\end{equation}
and therefore the photon flux should be $\Phi_{\gamma}=(1.2 \times 
10^{-4}-9.7 \times 
10^{-4})$~ph~cm$^{-2}$~s$^{-1}$. Basically, all the emitted photons are 
monochromatic with energy 
$E=m_s/2$. However, some of the photons will scatter with intersteller 
medium before reaching us. The probability of the scattering is 
\begin{equation}
P_s= \int_{\rm line~of~sight}n \sigma_cds,
\end{equation}
where $\sigma_c$ is the Compton cross section. For $n \approx 1$ 
cm$^{-3}$, the total $P_s$ from the disk and bulge is $\approx 1 \times 
10^{-2}$. Due to the scattering, the energy distribution of the photons 
reaching us is broadened slightly. Fig.~1 
shows the 
contribution of the emitted photon flux (we assumed $E=1$ MeV) together 
with the diffuse
background photon flux $dF/dE=2.62 (E/0.1~\rm 
MeV)^{-2.75}$~MeV$^{-1}$~cm$^{-2}$~s$^{-1}$ 
\citep{Kinzer}. The emitted photons contribute $2-15$ \% of the 
background flux at around 1 MeV which shows a small `MeV bump' in the 
spectrum. The MeV bump 
is a classical problem in observational astronomy which is long 
conjectured to be a real feature in the spectrum \citep{Kinzer}. However, 
the MeV bump is now commonly believed to be an artifact of incomplete 
background rejection 
\citep{Kinzer}. Nevertheless, at least in this model, part of the MeV 
bump is a real feature in the diffuse background photon spectrum.

\subsection{Active neutrino channel}
The lighter active neutrinos are produced in the main decay channel. 
The total active neutrino flux due to sterile neutrino decays passing 
through the earth is $0.02-0.1$~cm$^{-2}$~s$^{-1}$. The total number of 
active neutrinos passing through IceCube - the largest neutrino detector 
in the world - is about $10^9$~s$^{-1}$ 
\citep{Wiebusch}. 
Although this flux is theoretically detectable, the energy 
of the decayed neutrinos is too small to be 
detected in current experiments \citep{Lunardini}. 

Active neutrinos may 
interact with neutrons or protons in 
a pulsar to produce electrons or positrons. The cross 
section of such interactions is $\sigma_{\nu} \sim 
10^{-41}(E_{\nu}/10~{\rm MeV})^2$~cm$^2$, where $E_{\nu}$ is the energy of 
the neutrinos. For example, in a 
typical pulsar, the average number density is about $10^{38}$~cm$^{-3}$, 
and the 
mean free path for a 1 MeV neutrino in the pulsar is $10^5$ cm. 
As the crust thickness is also of order $10^5$ cm, almost every 
neutrino passing through a pulsar will interact with the neutrons and 
protons to 
produce electrons and positrons. As a result, a huge amount of electrons 
and positrons is produced and affected by the strong magnetic field $B$ 
in the pulsars to emit synchrotron radiation. The synchrotron 
frequency of the electrons is given by
\begin{equation}
f= \frac{\gamma_e^2 eB}{2 \pi m_ec}=2.8 \times 10^{18} \gamma_e^2 \left( 
\frac{B}{10^{12}~\rm G} \right)~ \rm Hz,
\end{equation} 
where $\gamma_e$ is the Lorentz factor of the electrons.
Therefore, the frequency of the synchrotron radiation lies within the 
x-ray band. The power emitted by one electron is 
\begin{equation}
P_{\rm syn}= \frac{2e^4B^2 \gamma_e^2}{3m_e^2c^3}= 7.9 \times 10^8 
\gamma_e^2 \left( \frac{B}{10^{12}~\rm G} \right)^2 ~\rm erg~s^{-1}.
\end{equation}
For a pulsar with radius 10 km nearby, 
the total number of active neutrinos due to sterile neutrino decays 
passing through the pulsar is $10^{12}$~s$^{-1}$. Assuming $B=10^{12}$ G 
and $\gamma_e =4$, the total power emitted 
is $\sim 10^{22}$~erg~s$^{-1}$, which is much less than the upper limit of 
the non-thermal x-ray luminosity in a typical pulsar 
($10^{30}$~erg~s$^{-1}$) \citep{Zavlin}. Therefore, only a very small peak 
near the 
synchrotron frequency may appear in the x-ray spectrum. Assuming that the 
$10^5$ or so pulsars in Milky Way all see similar neutrino flux, the 
resulting synchrotron radiation can contribute about 
$10^{27}$~erg~s$^{-1}$ to the background x-ray \citep{Lyne}. 

\section{Discussion and Summary}
The fact that active neutrinos have finite masses implies that 
right-handed 
neutrinos should exist which may indeed be massive sterile neutrinos. The 
existence of the sterile neutrinos has been invoked to explain many 
astrophysical phenomena such as the cooling flow problem in clusters 
\citep{Chan}. In this article, we consider the possibility that the decays 
of MeV sterile 
neutrinos act as a source of the 511 keV flux line. The decaying sterile 
neutrinos provide a diffuse source of positrons which can account for the 
required flux. The large bulge to disk ratio of 511 keV luminosity can 
also be accounted for if the decaying sterile neutrinos follow the 
isothermal distribution. From the observed 511 keV photon flux in the 
Milky Way, 
we obtain the allowed ranges of the sterile neutrino decay rate $\Gamma 
\ge 10^{-28}$~s$^{-1}$. Although 
we do not have representative tight bounds on decay rate of sterile 
neutrinos, the results are still compatible with 
cosmological bounds and cluster cooling flow ($\Gamma \le 
10^{-17}$~s$^{-1}$) \citep{Chan}. The radiative decay channel produces 
soft gamma rays, with an expected flux of $1.2 \times 
10^{-4}- 9.7 \times 10^{-4}$~ph~cm$^{-2}$~s$^{-1}$, which show up as a 
small MeV bump in the background photon spectrum. The total active 
neutrino flux due to sterile neutrino decays is estimated to be 
$0.02-0.1$~cm$^{-2}$~s$^{-1}$ in the vicinity of the earth. These active 
neutrinos interact 
with neutrons and protons in 
pulsars to produce x-ray photons which may be detectable in the 
future. 

\section{Acknowledgements}
This work is partially supported by a grant from the Research Grant
Council of the Hong Kong Special Administrative Region, China (Project No.
400805).

\bibliographystyle{spr-mp-nameyear-cnd}
\bibliography{biblio-u1}

\clearpage

\begin{figure*}
\vskip5mm
 \includegraphics[width=84mm]{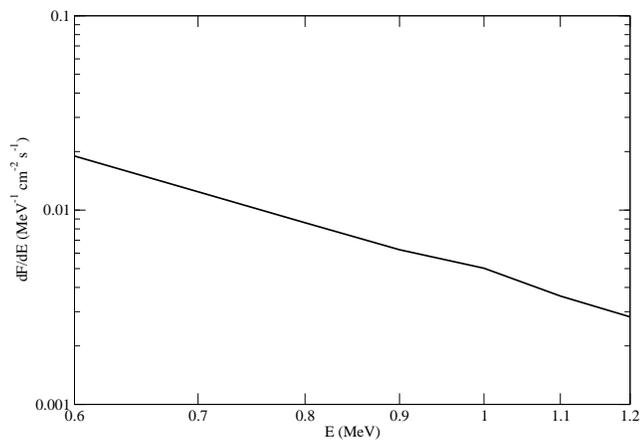}
 \caption{The spectrum of the background photons including the MeV 
photons coming from sterile neutrino decays.} 
\end{figure*}

\end{document}